\documentclass[12pt,preprint]{aastex}
\slugcomment{The Astrophysical Journal, 585:52-56, 2003 March 1}

\shorttitle{Constraints on Cardassian Expansion from High-z SNeIa}
\shortauthors{Zong-Hong Zhu and Masa-Katsu Fujimoto}

\begin{document}

\title{
	Constraints on Cardassian Expansion from Distant type Ia Supernovae
	}

\author{
	Zong-Hong Zhu
	and
	Masa-Katsu Fujimoto
	}

\affil{
	National Astronomical Observatory,
                2-21-1, Osawa, Mitaka, Tokyo 181-8588, Japan\\
	zong-hong.zhu@nao.ac.jp, 
	fujimoto.masa-katsu@nao.ac.jp
      }

\begin{abstract}
The distant type Ia supernovae data compiled by Perlmutter et al. (1999)
  are used to analyze the Cardassian expansion scenario, 
  which was recently proposed by Freese and Lewis (2002)
  as an alternative to a cosmological constant (or more generally a dark energy
  component) in explaining the currently accelerating universe. 
We show that the allowed intervals for $n$ and $z_{eq}$, the two parameters 
  of the Cardassian model, will give rise to a universe with a very low matter 
  density, which can hardly be reconciled with the current value derived from 
  the measurements of the cosmic microwave background anisotropy and galaxy 
  clusters (cluster baryon fraction).
As a result, this Cardassian expansion 
  proposal does not seem to survive the magnitude-redshift test for the present
  type Ia supernovae data, unless the universe contains primarily baryonic 
  matter. 
\end{abstract}


\section{Introduction}

A major development in modern cosmology is the discovery of the acceleration
  of the universe through observations of distant type Ia supernovae
  (Perlmutter et al. 1998, 1999; Riess et al. 1998, 2001; Leibundgut 2001).
It is well known that all known types of matter with positive pressure generate
  attractive forces and decelerate the expansion of the universe -- 
  conventionally, a
  deceleration factor is always used to describe the status of the universe's
  expansion \cite{san88}. 
Given this, the discovery from the high-redshift type Ia supernovae 
  indicates the existence of a new component with fairly negative
  pressure, which is now generally called dark energy, 
  such as a cosmological constant \cite{wei89,car92,kra95,ost95}
  or an evolving scalar field (referred to by some as quintessence) 
      \cite{rat88,wet88,fri95,cob97,cal98,gon02}. 
While current measurements of the cosmic microwave background anisotropies
  favor a spatially flat universe with cold dark matter \cite{ber00,lan01},
  both the deuterium abundance measured in four high-redshift hydrogen clouds
  seen in absorption against distant quasars (Burles and Tytler 1998a,b)
  (combined with the baryon fraction in galaxy clusters from X-ray data -- 
   see White et al. 1993 for the method)
  and the large-scale structure in the distribution of galaxies 
   \cite{bah00,pea01} have made a strong case for a low density universe 
  (for a recent summary, see Turner 2002a).
It seems that all these observations can be concordantly explained by the 
  hypothesis that there exists, in addition to cold dark matter, 
  a dark energy component with negative pressure in our universe \cite{tur02b}.
The existence of this component has also been independently indicated by
  other observations such as
  the angular size-redshift relations for compact radio sources
       (Guivits et al. 1999; Vishwakarma 2001; Lima and Alcaniz 2002; 
          Chen and Ratra 2003)
     and FRIIb radio galaxies
       (Guerra et al. 2000; Daly and Guerra 2002; Podariu et al. 2003),
  the age estimates of old high-redshift galaxies
    (Dunlop et al. 1996; Krauss 1997; Alcaniz and Lima 1999)
  and gravitational lensing 
    (Kochaneck 1996; Chiba and Yoshii 1999; Futamase and Hamana 1999;
      Jain et al. 2001; Dev et al. 2001; Ohyama et al. 2002; Sereno 2002).

Neither a cosmological constant nor a quintessence, the present
  candidates for the universe acceleration mechanism, 
  however avoid the cosmic coincidence problem --
  why the densities of dark energy and dark matter are comparable
  today (another related but distinct difficulty is the fine-tuning problem,
  see Carroll et al. 1992 for a discussion of this point).
Although the tracking field model \cite{zla99} provides a possible
  resolution to this problem, a convincing dark energy model with a solid
  basis in particle physics is still far off.
Therefore it is desirable to explore alternative possibilities, such as
  higher dimensions \cite{def02,gu02} or an altered theory of gravitation
  \cite{beh02}.
Very recently, Freese and Lewis (2002) proposed the ``Cardassian
  Expansion Scenario'' in which the standard Friedman-Robertson-Walker (FRW) 
  equation is modified as follows,
\begin{equation}
\label{eq:ansatz}
H^2 = A\rho + B\rho^n
\end{equation}
where $H \equiv \dot R / R$ is the Hubble parameter as a function of cosmic 
  time, $R$ is the scale factor of the universe and $\rho$ is the energy 
  density of matter and radiation.
In the usual FRW equation $B = 0$. To be consistent with the usual FRW
  result, one should take $A = 8\pi G/3$.
It is convenient to use the redshift $z_{eq}$, at which the two terms of
  eq.(\ref{eq:ansatz}) are equal, as the second parameter of the Cardassian
  model.
In this parameterization of ($n, z_{eq}$), it can be shown that 
  (Freese and Lewis 2002),
  $B=H_0^2(1+z_{eq})^{3(1-n)}\rho_0^{-n} [1+(1+z_{eq})^{3(1-n)}]^{-1}$,
  where $\rho_0$ is the matter density of the universe at the present time
  and $H_0=100h\,$kms$^{-1}$Mpc$^{-1}$ is the Hubble constant.
This particular proposal is very intriguing because the expansion of the
  universe will be accelerated automatically later without any dark energy 
  component -- the second term, which may arise as a consequence of brane 
  world cosmologies, dominates at a late epoch and drives the acceleration 
  of the universe.
It is valuable to explore the agreement of the Cardassian expansion model
  with the currently available cosmological observation data, as suggested 
  by Freese and Lewis, who proposed this scenario.
In a previous paper, the authors have used the recent measurements of the 
  angular size of high-redshift compact radio sources made by Gurvits et al. 
  (1999) to test the Cardassian model (Zhu and Fujimoto 2002).
It was shown that the allowed intervals for the two model parameters,
  $n$ and $z_{eq}$, are heavily dependent on the value of the mean
  projected linear size $l$ (see Table~1 of Zhu and Fujimoto 2002).
For example, at $l = 16h^{-1}$pc, the best fit occurs for
  $n=0.76$ and $z_{eq}=1.78$, which gives a reasonable matter density of
  $\Omega_m \sim 0.32$.
However, this analysis shows that, if one minimizes $\chi^{2}$ for the
  parameters $l$, $n$ and $z_{eq}$ simultaneously, the best fit to the current 
  angular size data prefers the conventional flat $\Lambda$CDM model to the 
  Cardassian expansion proposal. 
In this work, we analyze this scenario with the distant type Ia 
  supernovae sample compiled by Perlmutter et al. (1999).
It is shown that the allowed intervals for $n$ and $z_{eq}$ would give rise 
  to a universe with an unreasonably low matter density ($\Omega_m < 0.1$).
As a result, the Cardassian expansion scenario does not seem to survive the 
  magnitude-redshift test from the present type Ia supernovae data, 
  unless the universe contains no dark matter.
After providing a brief description of the data analysis method (section~2),
  we present our numerical results in section~3. Finally we summarize our
  conclusions and present discussion(section 4).

\section{Outline of the data analysis method}

The apparent bolometric magnitude $m(z)$ of a standard candle with absolute
  bolometric magnitude $M$ is related to the luminosity distance $d_L$ by
  $m=M+5\log d_L + 25$. 
Following Perlmutter et al. (1997), we write 
  the B-band magnitude-redshift relation as
\begin{equation}
\label{eq:mB}
m_B = {\cal M}_B + 5 \log {\cal D}_L,
\end{equation}
where ${\cal D}_L \equiv H_0 d_L$ is the ``Hubble-constant-free'' luminosity
  distance and ${\cal M}_B \equiv M_B - 5 \log H_0 + 25$ is the
  ``Hubble-constant-free'' $B$-band absolute magnitude at maximum of a 
  type Ia supernova.

In a Friedmann-Robertson-Walker (FRW) universe, the luminosity distance 
$d_L$ can be calculated from the redshift-dependent Hubble parameter,
   $H(z) = H_0 E(z)$, by the integral
$
d_L(z) = {(c / H_0) }{(1+z)} \int_{0}^{z} {dz^{\prime}
                                                / E(z)} .
$
For the ansatz of eq.(\ref{eq:ansatz}) and a flat universe with only matter,
   Freese and Lewis (2002) get
\begin{equation}
\label{eq:newE}
E^2(z; n, z_{eq}) = (1+(1+z_{eq})^{3(1-n)})^{-1}\times (1+z)^3
        + (1-(1+(1+z_{eq})^{3(1-n)})^{-1})\times (1+z)^{3n}
\end{equation}
where $n$ and $z_{eq}$ are the two paramters of the Cardassian model.

We use the Perlmutter et al. (1999) data to place observational constraint 
  on the Cardassian model parameters $n$ and $z_{eq}$.
This data set, plotted in Fig.~1, consists of 42 high-redshift Type Ia 
  supernovae from the 
  Supernova Cosmology Project\footnote{Supernova Cosmology Project:
  http://www-supernova.lbl.gov}, and 18 low-redshift Type Ia supernovae from 
  the Cal\'{a}n/Tololo Supernova Survey (Hamuy et al. 1996).
Both sets include corrections for the lightcurve width-luminosity relation.
The error bars, which include both the measurement errors and the intrinsic
  luminosity dispersion, have also been shown in Fig.~1.  
We determine the model parameters $n$ and $z_{eq}$ through a $\chi^{2}$ 
  minimization method.
The range of $n$ spans the interval [-3, 2] in steps of 0.01, while the
  range of $z_{eq}$ spans the interval [0, 4] also in steps of 0.01.
\begin{equation}
\label{eq:chi2}
\chi^{2}({\cal M}_B; n, z_{eq}) =
  \sum_{i}^{}{\frac{\left[m_B (z_{i}; {\cal M}_B; n, z_{eq}) 
     - m_{Bi}^{\rm eff}\right]^{2}}{\sigma_{m_{Bi}}^{2}}},
\end{equation}
where $m_B (z_{i}; {\cal M}_B; n, z_{eq})$ refers to the theoretical prediction
  from eq.(\ref{eq:mB}), $m_{Bi}^{\rm eff}$ is the observed effective magnitude,
  and $\sigma_{m_{Bi}}$ is the total uncertainty ($i$ refers to the $i$th
  supernova of the sample).
The summation is over all of the observational data points.

Evaluating the ansatz of eq.(\ref{eq:ansatz}) at the present time, 
  we have (Freese and Lewis 2002)
\begin{equation}
\label{eq:H0}
H_0^2 = {8 \pi G \over 3} \rho_0 [1+(1+z_{eq})^{3(1-n)}] .
\end{equation}
Because in the Cardassian model the universe is flat and contains only matter,
  the matter density at present, $\rho_0$, 
  should be equal to the `critical density' of this scenario.
From eq.(\ref{eq:H0}), we have
\begin{equation}
\label{eq:rhoc}
\rho_0 = \rho_{c,{\rm cardassian}} = \rho_{c} \times F(n), \;\;\;\;\;
  F(n) = [1+(1+z_{eq})^{3(1-n)}]^{-1}
\end{equation}
where $\rho_c = 3H_0^2/8\pi G$ is the critical density of the standard FRW 
  model.
Therefore the new critical density $\rho_{c,{\rm cardassian}}$ depends on the
  two parameters, $n$ and $z_{eq}$, 
  while $F(n, z_{eq}) \equiv \Omega_m$ gives the matter density in units of 
  the critical density of standard FRW model (Freese and Lewis 2002).
Instead of specifying $\Omega_m$ (or $F$), we consider both $n$ and $z_{eq}$ as
  independent paramters, while $\Omega_m(F)$ is treated as the output
  of the fitting result.
The magnitude ``zero point'' ${\cal M}_B$ can be determined from the 18 
  low-redshift
  supernovae that are carefully chosen from a sample of 29 supernovae from
  the Calan/Tololo survey.
After appropriate correction, they give
  ${\cal M}_B = -3.32\pm 0.05$ (Perlmutter et al. 1997).
We will use ${\cal M}_B = -3.27, -3.32, -3.37$ respectively as typical values
  of the zero point to fit the data, as well as a range of 
  ${\cal M}_B =-3.00\,$--$\,-3.60$ \cite{mes02} to check the robustness of
  our results.
In order to make the analysis independent of the choice of ${\cal M}_B$, we 
  also minimize eq.(\ref{eq:chi2}) for ${\cal M}_B$, $n$ and $z_{eq}$ 
  simutaneously, which we refer as the ``best fit''.

\section{Numerical results}

Table~1 summarizes our fitting results to the Cardassian expansion model.
Following Perlmutter et al. (1999), we analyze the 60 supernovae
  as three different sample groupings. 
Sample~A is the entire data set.
Sample~B excludes four outliers -- the two of them with lower redshifts,
  SN1992bo and SN1992bp, are the most significant outliers from the average
  light-curve width, while the other two with higher redshifts, SN1994H and
  SN1997O, are the largest residuals from $\chi^2$ fitting.
Sample~C further excludes two very likely reddened supernovae, SN1996cg and
  SN1996cn.
(For details of all these outliers, see Perlmutter et al. 1999.)
As shown in Table~1, the fitting results for samples A and B are very similar
  except for their goodness-of-fits.
The larger $\chi^2$ per degree of freedom (d.o.f) for sample~A, 
  $\chi_{\nu}^2 = 1.78$,
  indicate that the outlier supernovae included in this sample are probably not
  part of a Gaussian distribution and thus will not be appropriately weighted 
  in a $\chi^2$ fit (Perlmutter et al. 1999).
The $\chi^2$ per d.o.f for sample~B, $\chi_{\nu}^2 = 1.20$, is reduced 
  significantly and indicates that no large statistical errors remain 
  unaccounted for.
The fit for sample~C is a more robust one, because the two very likely reddened 
  supernovae, SN1996cg and SN1996cn, have been further removed 
  (Perlmutter et al. 1999).
All three best-fits result in the same value of the zero-point magnitude
  ${\cal M}_B=-3.42$, which is higher than ${\cal M}_B=-3.32$, the conclusion
  of Perlmutter et al. (1997,1999), but a little bit lower than
  ${\cal M}_B=-3.45$, the value of Efstathiou (1999) which is obtained from
  the best-fit to the combined data for type 
  Ia supernovae and the cosmic microwave background anisotropies.
The best fit to sample~C, with ${\cal M}_B=-3.42$, $n=-1.33$, $z_{eq}=0.43$
  and the lowest $\chi^2$ per d.o.f. of $1.11$, is 
  depicted in Fig.~1 as a solid line.
For comparison, three other curves with model parameters $n$ and $z_{eq}$
  taken from the Table~1 of Freese and Lewis (2002) are also shown.
The ability of these curves to fit the data is surprisingly distinct 
  (and can even be seen by eye),
  while the former solid curve matches the data points very well, 
    none of the later three curves does.

In Fig.~2, we show the confidence regions (68.3\% and 95.4\% C.L.) of 
  the fitting results in the plane ($n, z_{eq}$).
The three left panels show the results for sample A, B and C using the value of
  ${\cal M}_B = -3.32$ which was initiated by Perlmutter et al. (1997, 1999),
while the three right panels show their corresponding best fits 
  (i.e., the case of ${\cal M}_B = -3.42$).
In order to evaluate how reasonable the resulting parameters,
  $n$ and $z_{eq}$ are, two areas which give a currently optimistic matter
  density, $\Omega_m (F) = 0.330 \pm 0.035$, (see Turner 2002a for the argument)
  and a wider range of $\Omega_m (F) = 0.2$--$0.4$, respectively are also 
  shown in every panel.
As it is distinctly shown that there is no overlap between the resulting 
  parameter range and the reasonable area for the matter density,
  we have a 95.4\% confidence level (C.L.) in saying that the Cardassian 
  expansion is not compatible with a cold dark matter dominated universe with
  $\Omega_m (F) = 0.2$--$0.4$ 
  (the C.L. goes up to 99\% if the matter density of the universe is, 
  $\Omega_m (F) = 0.330 \pm 0.035$).
As a matter of fact, all of our fitting results for $n$ and $z_{eq}$, 
  point to a universe with $\Omega_m (F)$ less than 0.1, 
  which is unreasonable in light of the currently available
  cosmological observations \cite{bah00,pea01,tur02a}.

One could use the observational constraints on the deceleration factor $q$ to
  cross-check the robustness of the fitting results and the difficulty of the 
  Cardassian model with a reasonable matter density 
  (e.g., $\Omega_m \sim 0.33$) in explaining the type Ia supernovae data.
For the Cardassian expansion scenario parameterized by $n$ and $z_{eq}$, we
  get the deceleration parameter as a function of redshift
\begin{equation}
\label{eq:deceleration}
q(z) \equiv -{\ddot{R}R\over {\dot{R}}^2} 
     = -1 + {1 \over 2}{{\rm d}\ln E^2(z;n,z_{eq}) \over {\rm d}\ln (1+z)}
\end{equation}
where the function $E^2(z;n,z_{eq})$ is given by eq.(\ref{eq:newE}),
in which the $(1+z)^3$--dependent term dominates at high redshifts, causing the
  deceleration of the expansion of the universe.
An acceleration will only occur when it becomes negative, i.e. $q < 0$, 
  at a late epoch.
We plot the redshift dependent deceleration parameter in Fig.~3 for the 
  Cardassian models with the parameters of $n$ and $z_{eq}$ taken from the
  Table~1 of Freese and Lewis (2002) and from our best fits.
The shaded area in Fig.~3 corresponds to the present observational constraints
  on the deceleration parameter, i.e., the universe switched from deceleration
  to acceleration at a redshift interval $0.6 < z_{q=0} <1.7$ at the 1$\sigma$
  level (Perlmutter et al. 1999; Riess et al. 1998,2001; Avelino et al. 2001;
  Avelino and Martins 2002).
The problem now is apparent: while the best fits of this work predict the
  turnaround redshift well within the observation constraints, all Cardassian
  models with a reasonable matter density ($\Omega_m \sim 0.33$) (parameters 
  taken from the Table~1 of Freese and Lewis 2002) predict the turnaround 
  redshift less than $\sim 0.6$, 
  which is only marginally compatible to the present observations.
We are thrown into a dilemma: an unrealistically low matter density is needed 
  so that the acceleration starts early enough to be realistic (See Avelino
  and Martins 2002 for the same discussion for another brane world cosmology).
Therefore our fitting results are robust in explaining the type Ia supernovae 
  data of Perlmutter et al. (1999): all of them predict a universe with
  very low matter density ($\Omega_m < 0.1$).

\section{Conclusions and Discussion}

We have analyzed the Cardassian expansion recently proposed by Freese and 
  Lewis (2002) using distant type Ia supernovae data complied
  by perlmutter et al. (1999).
Although this particular proposal is an intriguing mechanism for the
  acceleration of the universe because it postulates the universe is flat, 
  matter dominated and accelerating, but contains no vacuum contribution, it is
  strongly disfavored by the present high-redshift type Ia supernovae data
  and the constraint of $\Omega_m \sim 0.3$.
The main point is that all fitting results of this scenario to the supernovae
  sample lead to a universe with unreasonably low matter density, leaving
  no space for the huge amount of dark matter whose existence has been widely
  accepted among the astronomical community 
  (see, e.g., Primack 2002, Turner 2002b).
Even if one can say that this Cardassian model can marginally pass the 
  cosmological test from the updated angular size data (Zhu and Fujimoto 2002),
  it can hardly survive the magnitude-redshift test for the present type Ia
  supernovae data unless the universe contains primarily baryonic matter.
There seems to be a tendency: a model that excludes the dark energy component
  dispels dark matter also (see Avelino and Martins 2002 for another analysis).
However, it is worth keeping in mind that a universe with low matter
  density $\Omega_m \sim 0.1$ can also fit the data of Perlmutter et al. (1999)
  surprisingly well (M\'esz\'aros 2002).

One of the major uncertainties in the present analysis comes from the errors
  of the magnitude ``zero point'' ${\cal M}_B$. 
There are several ways to overcome this problem. 
First of all,  one can analyzes the data over a large enough range 
  of ${\cal M}_B$ to include almost all of the possibilities,
  and then calculate the probability distribution for the model parameters
  by integrating over it (Perlmutter et al. 1997, 1999). 
However this is not needed for our purpose, because all our fits with 
  the values of ${\cal M}_B$
  from -3.00 to -3.60 lead to a universe with very low matter density.
Second, one could pin down the value of ${\cal M}_B$ through a larger sample
  low-redshift supernovae. 
Databases of nearby SNeIa are becoming unprecedentedly abundant
  (see, e.g., Li et al. 2001).
 In particular, the Nearby SN Factory\footnote{Nearby SN Factory: 
  http://snfactory.lbl.gov} will accumulate a sample of 300 low-redshift 
  supernovae and determine  ${\cal M}_B$ to a precision of $\pm (0.01$--$0.02)$.
Finally, from eq.~(\ref{eq:mB}), the model parameters $n$ and $z_{eq}$ can be 
  determined by measuring differences of magnitudes at different redshifts, 
  which are independent of ${\cal M}_B$ \cite{fri02}.

Other uncertainties of cosmological parameter extraction from high-redshift 
  type Ia 
  supernovae sample caused by progenitor and metallicity evolution, extinction,
  sample selection bias, local perturbations in the expansion rate,
  gravitational lensing and sample contamination have been carefully studied
  by Riess et al. (1998) and Perlmutter et al. (1999). 
It was found that none of these effects can seriously change the result.
Considering our results show that the matter density predicted by the 
  Cardassian scenario is less than 0.1, it is impossible for any of these 
  effects to change the case. 
In short, the present Cardassian expansion model is strongly disfavoured by 
  the current distant SNeIa data and the constraint of $\Omega_m \sim 0.3$.
We hope that other convincing mechanisms for the acceleration of the universe
  will appear in the near future.

\acknowledgements

%%%
We would like to thank
  A. G. Riess and W. Li for their help on parameter extraction with 
    high-redshift type Ia supernovae data,
  Y. Tsunesada and S. Sato for their aasistance on using the computer cluster 
    ATAMA,
  and P. Beyersdorf for polishing up the English.
%%%
Z.-H. Zhu is also grateful to all TAMA300 member and the
        staffs of NAOJ for their hospitality and help during his stay.
%%%
Finally, our thanks go to the anonymous referee for valuable comments and
  useful suggestions, which improved this work very much.
%%%
This work was supported by
  a Grant-in-Aid for Scientific Research on Priority Areas (No.14047219) from
  the Ministry of Education, Culture, Sports, Science and Technology.

\clearpage

\begin{figure}
\plotone{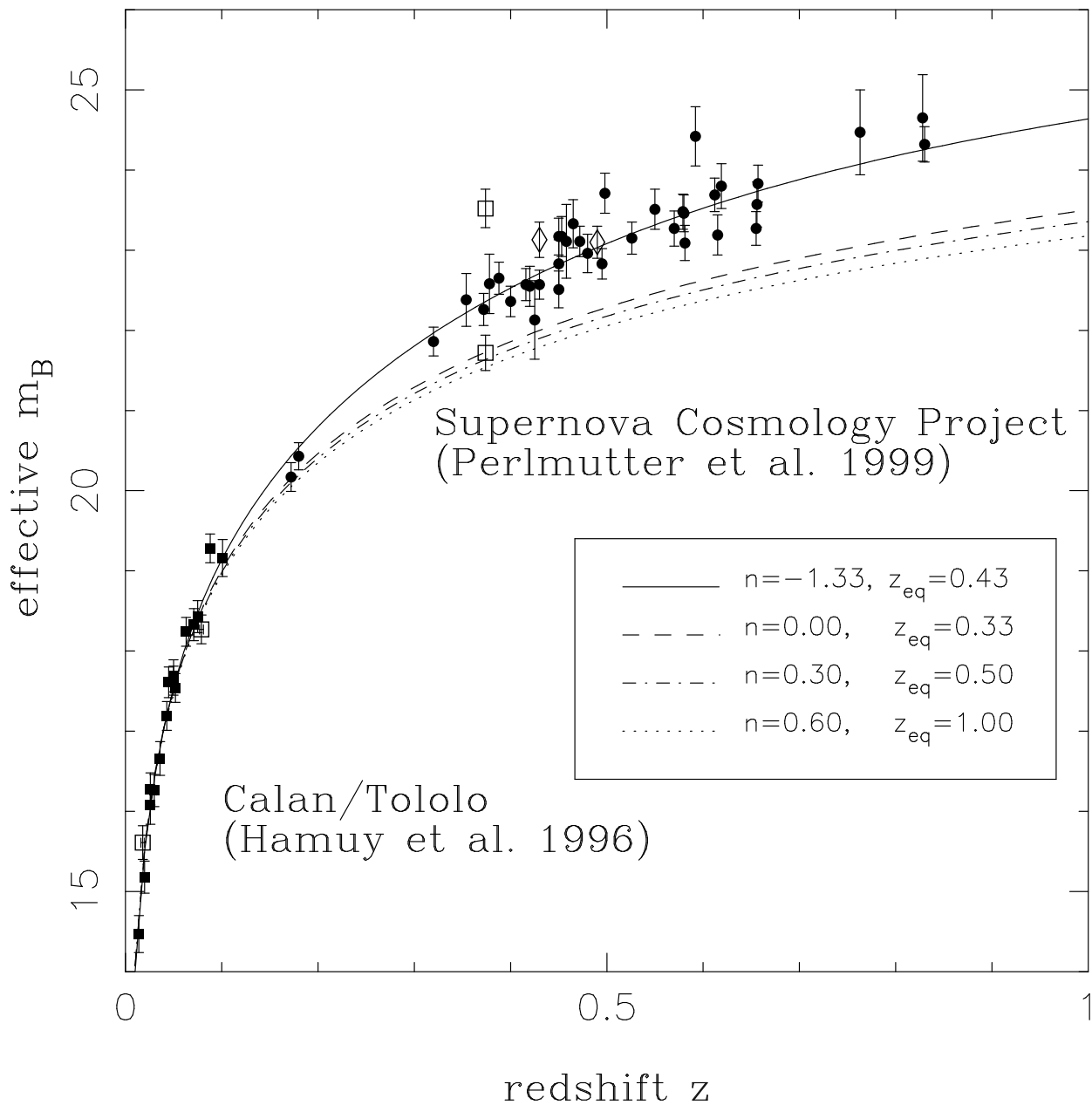}
\figcaption{Hubble diagram for 42 high-redshift SNeIa from the Supernova
	Cosmology Project and 18 low-redshift SNeIa from the Cala\'n/Tololo 
	Supernova Survey.
	The empty squares mark the four outliers which are excluded in
	sample B, while the empty diamonds mark the another
	two futher excluded in sample C.
	The solid curve corresponds to our best fit to sample C, 
	with ${\cal M}_B=-3.42$, $n=-1.33$, $z_{eq}=0.43$.
	The values of ($n$, $z_{eq}$) for the other three curves are taken
        from the Table~1 of Freese and Lewis (2002).
	\label{fig:data}
	}
\end{figure}

\clearpage

\begin{figure}
%\vspace*{-25.0mm}
\plotone{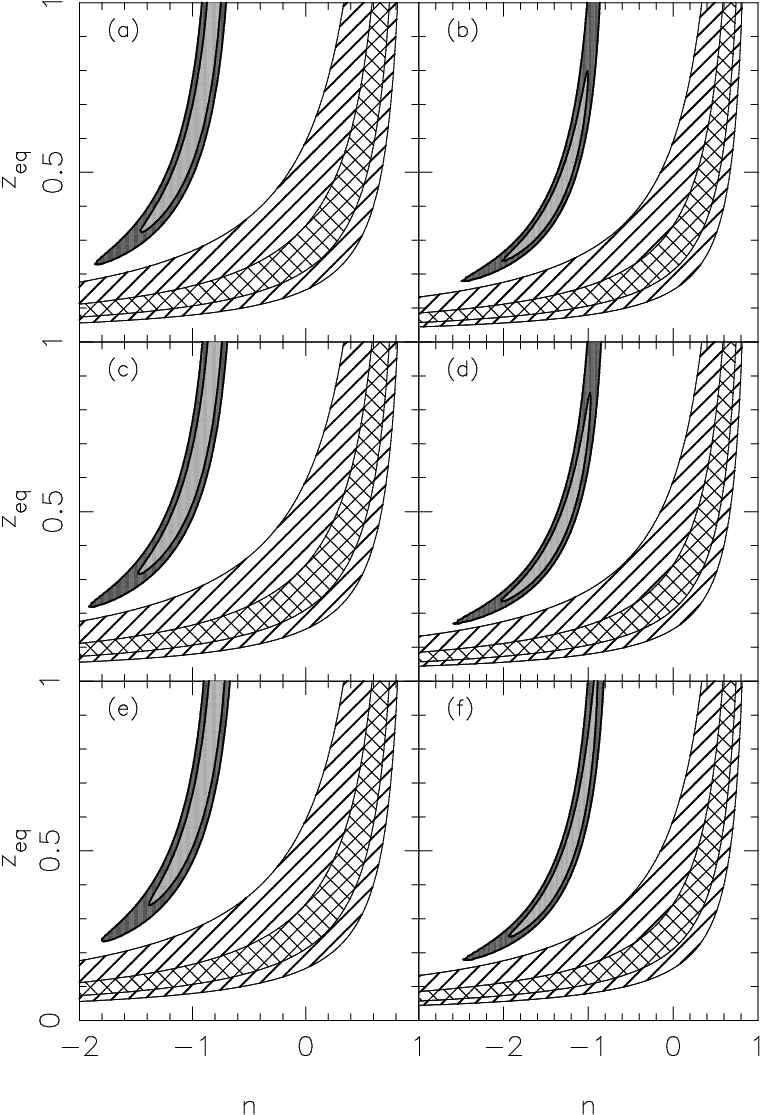}
%\figcaption{Confidence regions of fitting results of sample~A (a,b), 
%        B (c,d) and C (e,f), and the area constrained by observed matter
%        density of the universe in the $n\,$--$\,z_{eq}$ plane.
%        The faint(dark)-shaded areas show the parameter regions with
%        a confidence level of 68.3\% (95.4\%).
%        The positively slanted hatchings and the cross-hatched regions
%        correspond to the parameter areas that give the matter density
%        of the universe $\Omega_m$ ($F$)$=0.2$-- $0.4$ and
%                         $\Omega_m$ ($F$)$=0.330 \pm 0.035$ respectively.
%        The left three panels are for the fitting results with the assumption
%        of ${\cal M}_B=-3.32$,
%	 while the right three panels show the case of ${\cal M}_B = -3.42$.
%        \label{fig:contours}
%        }
\end{figure}
%\addtocounter{figure}{-1}
\begin{figure}
\figcaption{Confidence regions of fitting results of sample~A (a,b), 
	B (c,d) and C (e,f), and the area constrained by observed matter 
	density of the universe in the $n\,$--$\,z_{eq}$ plane.
	The faint(dark)-shaded areas show the parameter regions with
	a confidence level of 68.3\% (95.4\%).
	The positively slanted hatchings and the cross-hatched regions
	correspond to the parameter areas that give the matter density 
	of the universe $\Omega_m$ ($F$)$=0.2$-- $0.4$ and
			 $\Omega_m$ ($F$)$=0.330 \pm 0.035$ respectively.
	The left three panels are for the fitting results with the assumption
	of ${\cal M}_B=-3.32$,
	while the right three panels show the case of ${\cal M}_B = -3.42$.
        \label{fig:contours}
        }
\end{figure}

\clearpage

\begin{figure}
\plotone{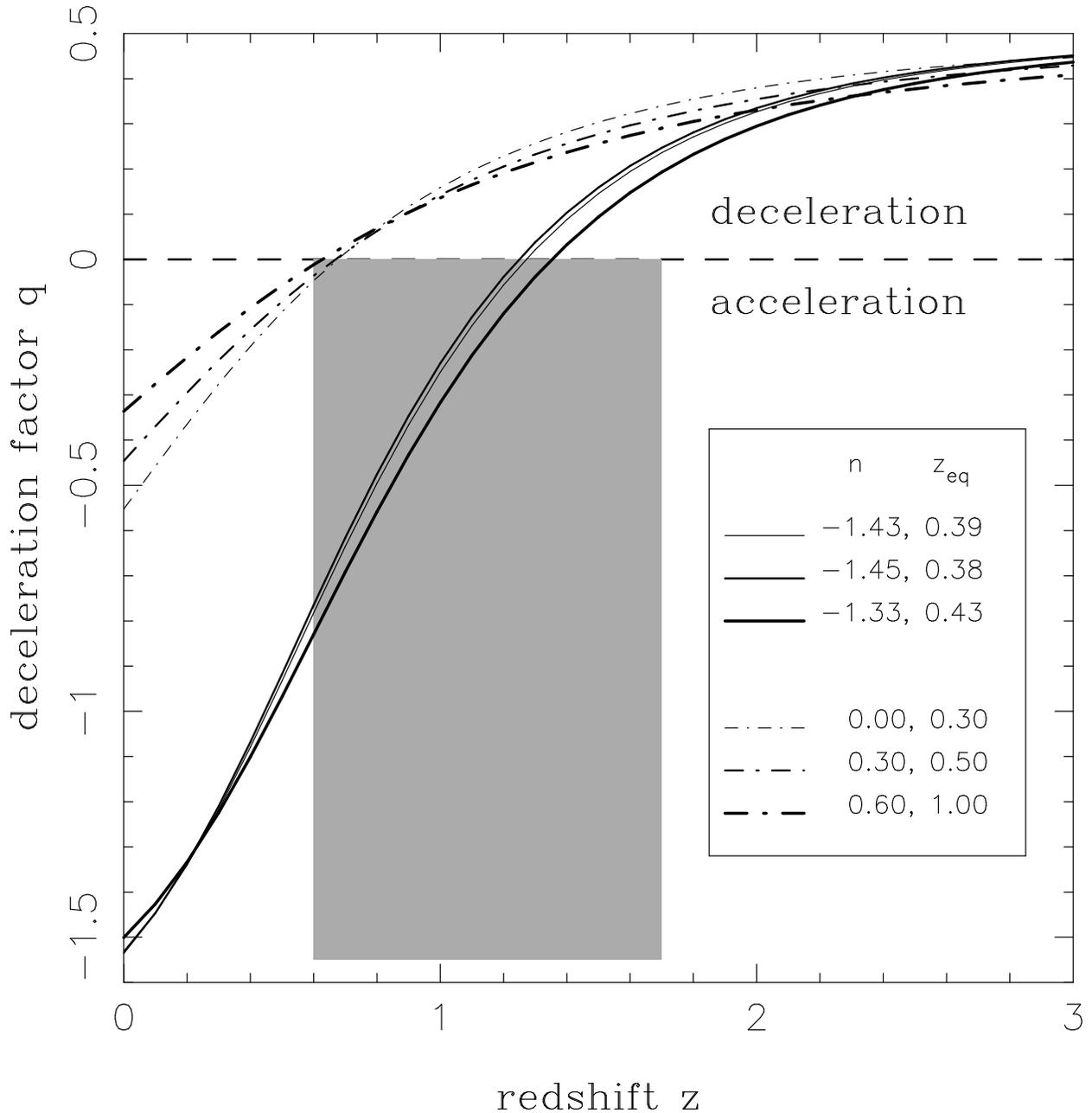}
\figcaption{Diagram of the deceleration parameter $q$ versus redshift for 
	the Cardassian expansion models.
	The model parameters are taken from Table~1 of Freese and Lewis (2002)
	and from our best fits respectively.
	The shaded area shows the observational constraint on $q$ from the
	literatures.
	As it shows, our best-fits are much more compatible with the
	observational constraints on the turnaround redshifts than the 
	Cardassian models with $\Omega_m (F) = 0.33$ are.
	}
        \label{fig:deceleration}
\end{figure}

\clearpage

\begin{deluxetable}{lcrcccc}
\tablecaption{
        Fitting results for the Cardassian model from distant type Ia 
	supernovae data compiled by Perlmutter et al. (1999).
        \label{tab:bestfits}
        }
\tablewidth{0pt}
\tablehead{
	\colhead{Sample} &
	\colhead{N} &
        \colhead{${\cal M}_B$}&
        \colhead{$n$}&
        \colhead{$z_{eq}$}&
        \colhead{$\Omega_m (F)$}&
        \colhead{$\chi^{2}$}
        }
\startdata
        A ......& 60&           -3.27& -0.81& 0.82& 0.037& 108.0\nl
        A ......& 60&           -3.32& -0.99& 0.62& 0.053& 104.3\nl
        A ......& 60&           -3.37& -1.19& 0.49& 0.068& 102.1\nl
        A ......& 60& Best fit: -3.42& -1.43& 0.39& 0.083& 101.3\nl
	B ......& 56&           -3.27& -0.80& 0.83& 0.037&  69.7\nl
        B ......& 56&           -3.32& -0.99& 0.61& 0.055&  66.4\nl
        B ......& 56&           -3.37& -1.21& 0.47& 0.072&  64.5\nl
        B ......& 56& Best fit: -3.42& -1.45& 0.38& 0.086&  63.9\nl
	C ......& 54&           -3.27& -0.71& 1.08& 0.023&  65.4\nl
        C ......& 54&           -3.32& -0.90& 0.72& 0.043&  62.4\nl
        C ......& 54&           -3.37& -1.12& 0.53& 0.063&  60.7\nl
        C ......& 54& Best fit: -3.42& -1.33& 0.43& 0.076&  60.2\nl
\enddata
\tablecomments{
  Sample A: all supernovae;     
  Sample B: excludes four outliers, SN1992bo, SN1992bp, SN1994H and SN1997O;
  Sample C: further excludes SN1996cg and SN1996cn.
	}
\end{deluxetable}

\end{document}